\documentclass[twocolumn]{aastex631}

\begin{document}
\title{The Kraft Break Sharply Divides Low Mass and Intermediate Mass Stars}

\author[0000-0002-0786-7307]{Alexa C. Beyer}
\affiliation{Department of Physics and Astronomy,
Georgia State University, 
Atlanta, GA 30303, USA}
\affiliation{Department of Physics, Astronomy, Geology and Environmental Science,
Youngstown State University,
Youngstown, OH 44555, USA}

\author[0000-0001-5313-7498]{Russel J. White }
\affiliation{Department of Physics and Astronomy,
Georgia State University, 
Atlanta, GA 30303, USA}

\begin{abstract}
Main sequence stars transition at mid-F spectral 
types from slowly rotating (cooler stars) to rapidly
rotating (hotter stars), a transition known as the Kraft Break
\citep{Kraft1967} and attributed the disappearance of the outer
convective envelope, causing magnetic braking to become ineffective.
To define this Break more precisely,
we assembled spectroscopic measurements of 405 F stars within
33.33 pc.  Once young, evolved and candidate
binary stars are removed, the distribution of projected rotational
velocities shows the Break to be well-defined and
relatively sharp.  Nearly all
stars redder than $G_{BP}-G_{RP}$ = 0.60 mag are slowly rotating
($v$sin$i$ $\lesssim$ 20 km/s), while only 4 of 40 stars bluer than
$G_{BP}-G_{RP}$ = 0.54 mag are slowly rotating, consistent with
that expected for a random distribution of inclinations.  The
Break is centered at an effective temperature of 6550 K and has a
width of about 200 K, corresponding to a mass range of $1.32 - 
1.41$ M$_\odot$.  The Break is $\sim$ 450 K hotter than the 
stellar temperature at which hot Jupiters show a change in their
obliquity distribution, often attributed to tidal
realignment.  The Break, as defined above, is nearly but
not fully established in the $\sim$ 650 Myr Hyades cluster;
it should be established in populations older than
1 Gyr.  We propose
that the Kraft Break provides a more useful
division, for both professional and pedagogical purposes, between
what are called low mass stars and intermediate mass stars; the
Kraft Break is observationally well-defined and is
linked to a change in stellar structure.

\end{abstract}

\section{Introduction} \label{sec:intro}

\citet{Kraft1967} demonstrated that nearby, non-emission line,
main sequence stars exhibit a change in their rotation rates near
spectral type F5.  Cooler stars are slowly rotating, often with
projected rotation rates below what can be measured with
spectrographs (e.g., $v$sin$i$ $\lesssim 10$ km/s), while hotter
stars rotate many times faster than this.
This change in rotation had been identified in previous work
\citep[e.g.,][]{AbtHunter1962, kraft1965}, but the
transition near F5 is now commonly referred to as
the ``Kraft Break."  Main sequence relations at the time \citep{iben1967} 
suggested the transition occurs at a stellar mass of about 1.2 M$_\odot$.

The dramatic change in rotation rate is not believed to be primordial
since almost all stars over this mass range have similar rotation
rates while young, at least up to about 1.6 M$_\odot$
\citep{gray1982, wolff1997}.  Instead, the change is attributed to
the effectiveness of magnetic braking.  A low mass star like the Sun
has an outer convective zone capable of generating a dynamo-like
magnetic field \citep{charbonneau2014}.  This field couples to 
the ionized stellar wind and magnetically brakes the star over time 
\citep[e.g.,][]{schatzman1962, weberdavis1967, matt2012, gallet_bouvier2013,
vansaders2019}.
In contrast, a slightly higher mass star, sometimes referred to as
an intermediate mass star \citep{aerts2019}, has an outer radiative
zone and is not expected to generate a magnetic field.  Since
these slightly higher mass stars age with no magnetic torque, they
retain the
angular momentum imparted during their pre-main sequence phase.

The location of the Kraft Break has provided an important
reference point for understanding how the angular momentum
evolution of stars relates to their internal structure
\citep[e.g.,][]{amard2016}.  However, determining the precise
spectral type / color / stellar mass where this Break occurs
and determining how sharp the Break is has been compromised by
subsets of low mass stars that remain rapidly rotating.  Since
magnetic braking takes 10s to 100s of millions of years to slow
the rotation rate \citep[e.g.,][]{bouvier2014}, young low mass are often
rapidly rotating.  Similarly, low mass short-period binary stars
remain rapidly rotating if they are tidally locked to their
companion \citep{zahnbouchet1989}.  And finally, higher mass,
post-main sequence stars that have evolved into cooler subgiant
stars can be rapidly rotating \citep{wilson_skumanich1964};
$\beta$ Cassiopeiae is a well-studied example of the latter
\citep{che2011, zwintz2020}.

Pioneering studies of stellar rotation relied mostly on
spectroscopy to measure the projected rotational velocity
($v$sin$i$), but this provides only a lower limit on the
equatorial rotational velocity.  Rotationally modulated
brightness variations caused by the presence of cool star
spots provides a more direct way of measuring the rotation
rate of a star, without the sin$i$ ambiguity.  Photometric
surveys of stars enabled
by the Kepler \citep{borucki2010}, K2 \citep{howell2014} and 
TESS \citep{ricker2014} missions have yielded
more than 100 thousand rotation periods for low and some
intermediate mass stars \citep[e.g.,][]{nielson2013,
mcquillan2014, koukel2022, reinhold2023, fetherolf2023,
phillips2024}.  These studies confirm a clear change in the
rotational distribution of stars near spectral type
F5\footnote{To our knowledge these studies haven't measured
the Kraft Break to be at spectral type F5, but are quoting F5
as originally claimed by \citet{Kraft1967} and reasonably consistent
    with their data.}.  However, because these studies include
distant, less well-studied stars, the Break is blurred by 
populations of young, binary and evolved stars.  Additionally,
the disappearance of the convective zone is expected to diminish
the presence of cool star spots, making it harder to measure
rotation rates for stars above the Kraft Break.  This is
confirmed by the steep decline in the fraction of measured
rotation periods with temperatures above 6000 K 
\citep[e.g.,][]{mcquillan2014}.  Projected rotational velocity
measurements (e.g., $v$sin$i$ values) of a large sample of stars
remain the least biased
way to map the rotation rate of stars across the Kraft Break.

In Section \ref{sec:sample} we use \textit{Gaia} data to assemble
a volume limited sample of F stars with $v$sin$i$ measurements.
In Section \ref{sec:exclude} we exclude young, candidate binary
and evolved stars from this sample.  In Section \ref{sec:break} we
define the location and width of the Kraft Break, and in Section
\ref{sec:hyades} we investigate how well it is established in the
Hyades open cluster.  In Section \ref{sec:discussion} we discuss
implications, including a suggestion to use the Kraft Break to
more formally divide low mass stars and intermediate mass stars.

\section{Sample Selection and Stellar Property Measurements}
\label{sec:sample}

To identify the Kraft Break more clearly, we use Gaia Data
Release 3 \citep[Gaia DR3;][]{gaiamission2016, gaiaDR32023}
to assemble all F stars within 33.33 pc of the Sun.  Specifically,
we query stars that have a parallax greater than 30 mas, a parallax
uncertainty less than 1.0 mas, and have a $G_{BP}-G_{RP}$ color
between 0.327 mag (spectral type A9) and 0.784 mag (spectral type
G0), according to \citet{PecautMamajek2013}.  We removed stars
fainter than G = 10.0 mag (190 stars) to exclude white dwarfs in
this color range.  The specific ADQL query used is:
\begin{verbatim}
select * from gaiadr3.gaia_source gs 
where gs.parallax > 30
and gs.parallax_error < 1.0
and gs.bp_rp > 0.327 
and gs.bp_rp <0.784
and gs.phot_g_mean_mag < 10.0
\end{verbatim}

This yields a sample of 426 stars.  We note
that not imposing a cut on the parallax uncertainty only increased
the sample by four stars; these nearby bright stars
(2.27 $<$ G $<$ 7.59) have well determined parallax
measurements.

\begin{figure}[ht!]
\centering
\plotone{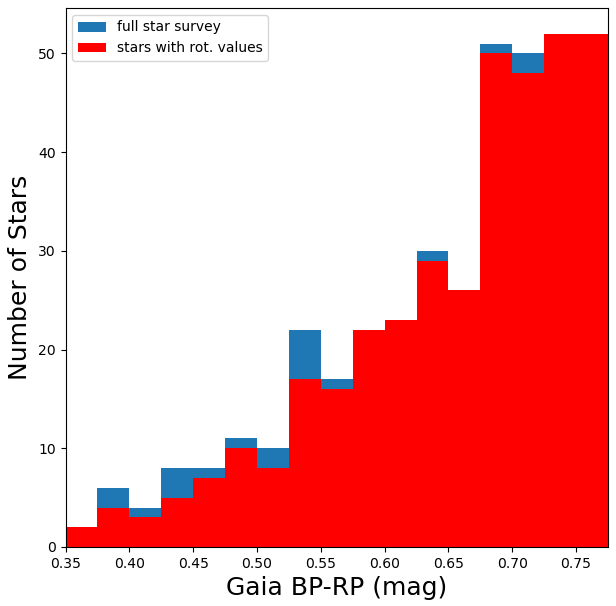}
\caption{Shown is the $G_{BP}-G_{RP}$ color distribution of 426
F stars within 33.33 pc of the Sun (\textit{blue bars}).  The
subset of these with projected rotational velocity
($v$sin$i$) measurements is indicated (\textit{red bars}).
\label{fig:color_distribution}}
\end{figure}

Of these 426 stars, Gaia DR3 provides spectroscopic measures of
effective temperature, surface gravity, metallicity and line
broadening measurements for 349 stars \citep{fouesneau2023},
although only 256 have Gaia line broadening measurements.
\citet{fremat2023} show that Gaia's line broadening measurement
(vbroad) agrees well with projected rotational velocity values
($v$sin$i$) over this temperature and visual brightness range.
We supplement these measurements with data from the
Geneva-Copenhagen survey targeting nearby F and G dwarfs
\citep{nordstrom2004}, and \citet{schroder2009} targeting
rapidly rotating Sun-like stars.  These added 135 and 14
additional stars, respectively, for a total of 405 stars
with line broadening measurements.  We caution that
\citet{nordstrom2004} report $v$sin$i$ values as low as 0 km/s,
which are likely unrealistic, and they round $v$sin$i$ values
to the nearest km/s below 30 km/s, or the nearest 5 or 10 km/s
above 30 km/s.  The $G_{BP}-G_{RP}$ color distributions of all
426 stars and the subset of 405 with spectroscopic measurements
are shown in Figure \ref{fig:color_distribution}.  

The distance limit of 33.33 pc is chosen to avoid reddening and to
maximize the fraction of stars with spectroscopic measurements and
accurate stellar property assessments (Section
\ref{sec:exclude}).

\begin{figure*}[ht]
\includegraphics[width=0.8\textwidth]{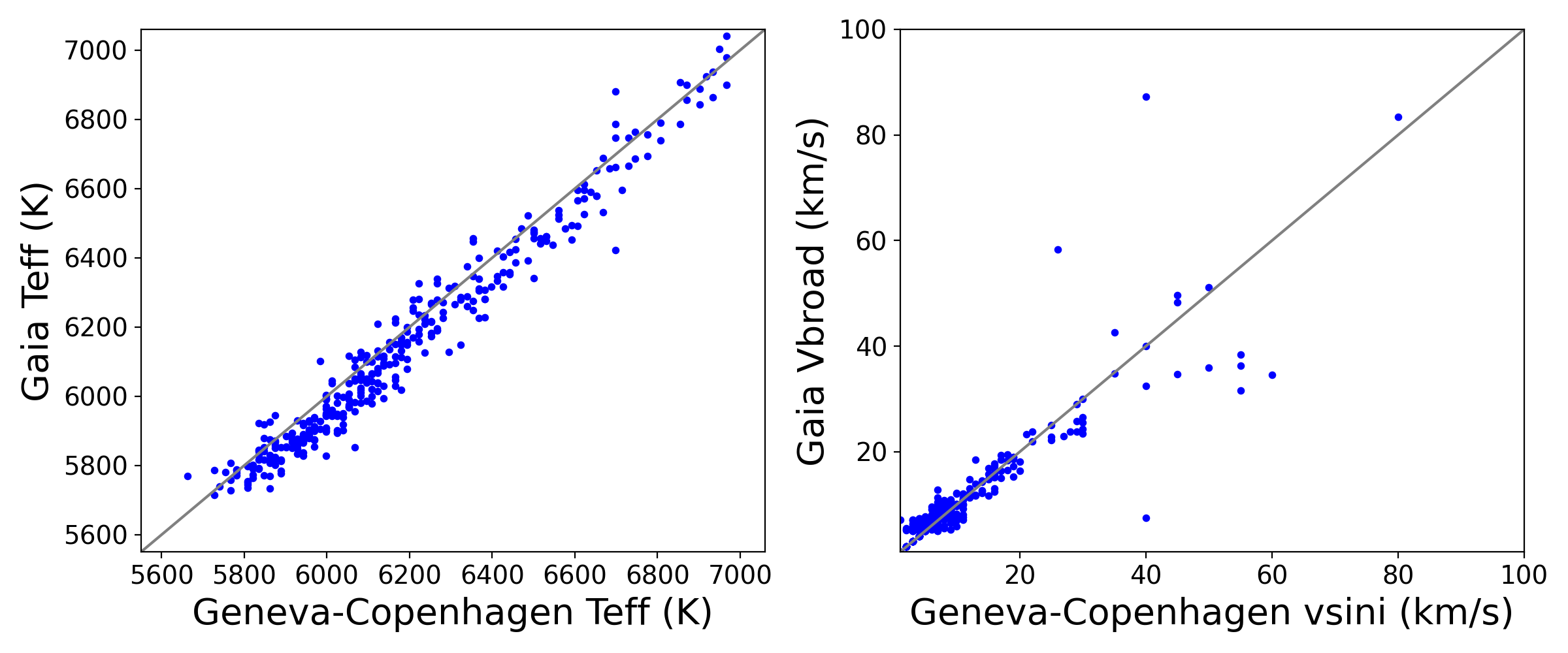}
\centering
\caption{Gaia DR3 measurements of effective temperature (\textit{left panel})
and projected rotational velocity (\textit{right panel}) are compared with
the same measurements from the Geneva-Copenhagen survey.}
\label{fig:comparisons}
\end{figure*}

To check for possible biases in the assembled measurements, we compare
spectroscopic measurements from Geneva-Copenhagen with those from Gaia
DR3 (Figure \ref{fig:comparisons}).  Over the temperature range of 5700 
K to 7000 K, a comparison of 337 temperature values shows that the 
Geneva-Copenhagen temperatures are on average hotter by 41 K, with the
standard deviation of differences being 56 K.  A comparison of 205 $v$sin$i$
values shows relatively good agreement up to $\sim 35$  km/s, with an
average difference of 0.2 km/s and a standard deviation of differences of
6.1 km/s.  Above 35 km/s, the measurements agree less well, with an
average difference of 2.0 km/s and standard deviation of differences
being 19.75 km/s.  However, the large dispersion is dominated by just 
3 stars (see Figure \ref{fig:comparisons}).

\begin{figure*} [ht]
\centering
    \includegraphics[scale=0.4]{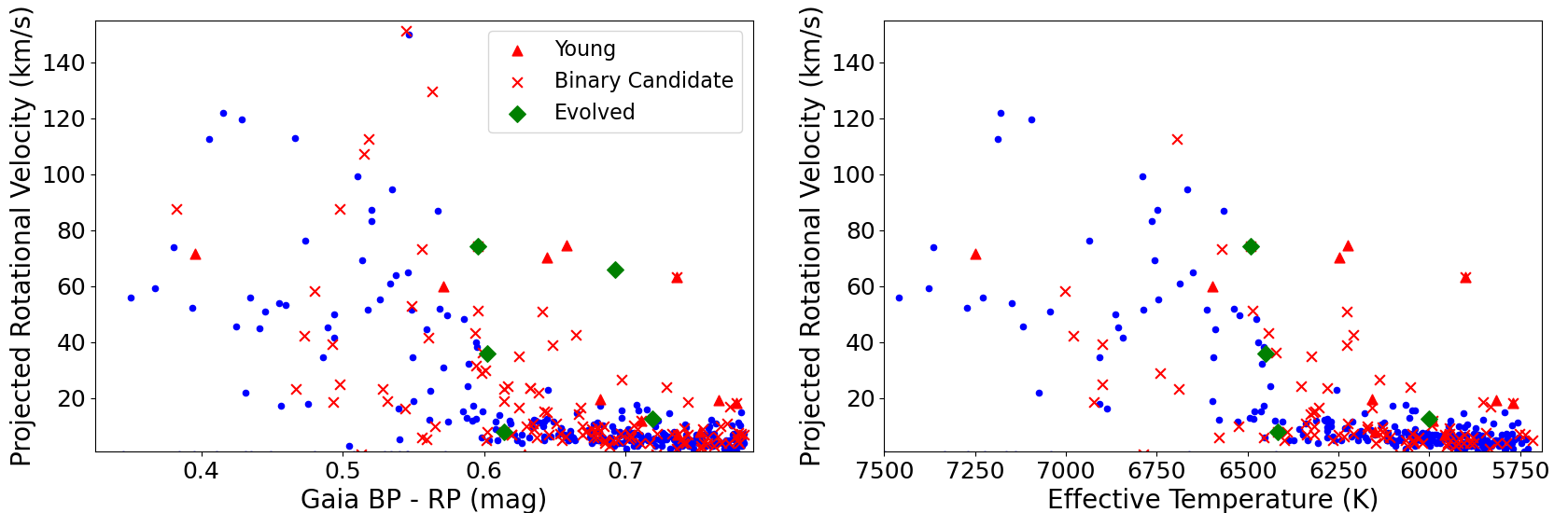}
    \caption{Shown are projected rotational velocities versus
    $G_{BP}-G_{RP}$ color (\textit{left panel}) and effective
    temperature (\textit{right panel}).  Stars identified as
    young, evolved or candidate binary are marked.  
    \label{fig:kraftbreak_all}}
\end{figure*}

\subsection{Excluding Young, Evolved and Candidate Binary Stars}
\label{sec:exclude}

The sample of 426 F stars within 33.33 pc
of the Sun is next vetted of stars that may have anomalous
rotation rates, as described in Section \ref{sec:intro}.  We
exclude all stars that are known members of nearby Moving Groups
with ages less than 200 Myr.  The 12 excluded stars include the
$\beta$ Pictoris Moving Group stars HD 29391 and HD 35850, the
Carina Association star HD 62848, the Columba Association star
HD 29329, and the AB Dor Moving Group stars HD 17332 A, HD 25457,
HD 38393, HD 45270, HD 139664, HD 141891 and HD 147584, HD 189245
\citep{nakajima2012, bell2015, gagne2018}.  The stars have ages
spanning from 24 Myr to 149 Myr \citep{bell2015} and projected
rotational velocities spanning from 7 to 74 km/s.  We note that
the remaining sample includes 4 members of the Ursa Majoris Moving
Group \citep[HD 91480, HD 113139 A, HD 115043 and HD 
116656;][]{gagne2018}, with an estimated age of $414\pm23$ Myr
\citep{jones2015}.  We exclude all 117 stars flagged as known or
candidate binaries in the Geneva-Copenhagen survey
\citep{nordstrom2004}.  Although we
only expect the stellar rotation to be tidally altered in short-period
systems, a stellar companion may nevertheless influence a star's
rotational
evolution \citep[e.g.,][]{patience2002, meibom2007} and the measure of
stellar rotation and other stellar parameters can be biased by the
presence of a spatially unresolved companion.  
We therefore conservatively exclude
all stars flagged as such.  In addition to these, we exclude stars
with spectroscopically measured surface gravity values (e.g.,
log\,$g$) less than 3.75 dex, as reported in Gaia DR3.  These are
suspected to be more massive stars that have evolved off the main
sequence and are transitioning through the F spectral type range
as subgiants.  The 5 excluded stars include HD 48737, HD 68456,
HD 198084, HD 219571, HD 224617; their projected rotational
velocities span from 8 to 74 km/s.  The sample of 405 F stars with
$v$sin$i$ measurements,
including those identified as young stars, candidate binary stars
and evolved stars is illustrate in Figure \ref{fig:kraftbreak_all}.  

\section{Defining the Kraft Break} \label{sec:break}

\subsection{Color and Effective Temperature}

\begin{figure*} [ht]
\centering
    \includegraphics[scale=0.4]{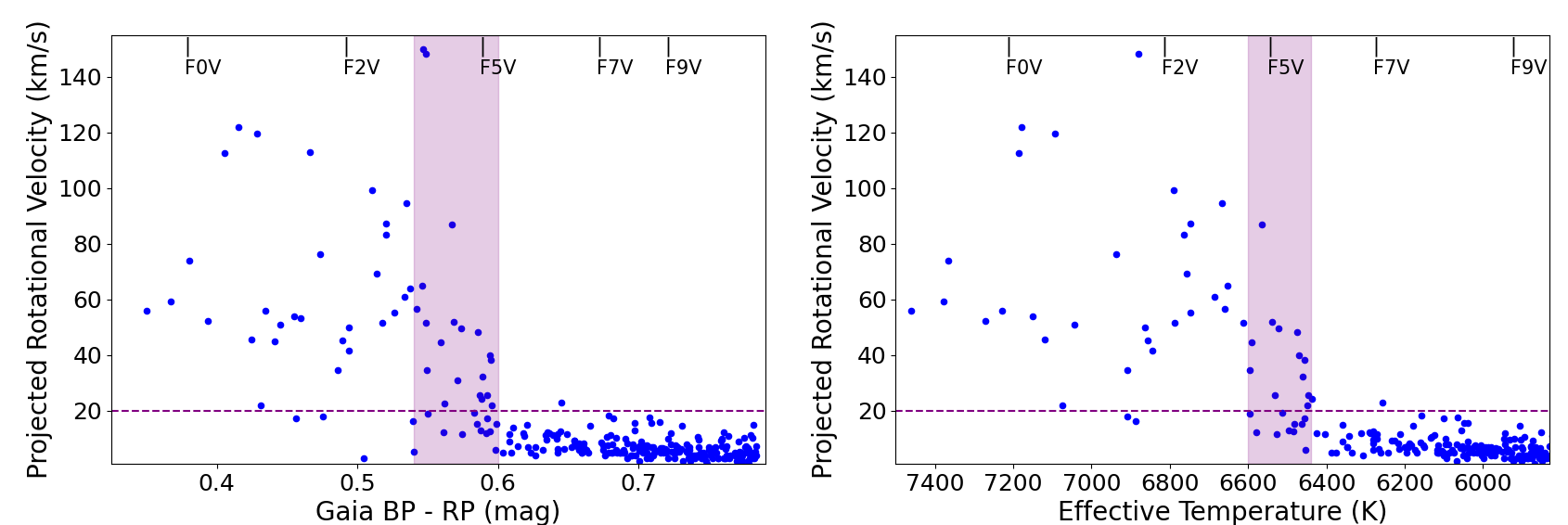}
    \caption{Shown are the projected rotational velocities versus 
    $G_{BP}-G_{RP}$ color (\textit{left panel}) and effective
    temperature (\textit{right panel}) after removing young,
    evolved and candidate binary stars.  The roughly uniform distributions
    of $v$sin$i$ values in the shaded regions are used to define the 
    location and width of the Kraft Break (see Section \ref{sec:break}).
    \label{fig:kraftbreak_clean}}
\end{figure*}

After removing young, evolved and candidate binary stars, 295 remain in
the sample.  These are considered mature, single, main sequence F stars,
278 of which have $v$sin$i$ measurements.  Figure 
\ref{fig:kraftbreak_clean} shows the distributions of projected
rotation velocity ($v$sin$i$) versus Gaia $G_{BP}-G_{RP}$ color and
versus effective temperature.  Both plots show an abrupt change in the
distribution of $v$sin$i$ values at mid-F spectral
types.  We define the Kraft Break to be the region with a roughly
uniform distribution of $v$sin$i$ values, above and below which
the distributions are distinct as confirmed by Kolmogorov–Smirnov
tests at the $>99.9$\% level.  In $G_{BP}-G_{RP}$ color, the Break spans
from 0.54 to 0.60.  These color limits correspond closest to the
$G_{BP}-G_{RP}$ colors of an F4V star (0.546 mag) and an F5V star
(0.587 mag), using the relations of \citet{PecautMamajek2013}.
In Figure \ref{fig:kraftbreak_clean}, the 217 stars redder than the
Break have relatively low $v$sin$i$ values,
with a mean value of 6.7 km/s and a standard deviation of 3.6 km/s.
In contrast, the 40 stars above the Break have a mean
$v$sin$i$ value of 63 km/s, and a standard deviation of 35 km/s.
We note that only 4 out of 40 stars above the Break (10 \%) have
$v$sin$i$ values less than 30 km/s, which is consistent with that
expected for a random distribution of inclinations.  The results
indicate that all mature, single, main sequence stars bluer than
$G_{BP}-G_{RP}$ = 0.54 are rapidly rotating ($v$sin$i$ $\gtrsim$
20 km/s) while those redder than $G_{BP}-G_{RP}$ = 0.60 are
slowly rotating ($v$sin$i$ $\lesssim$ 20 km/s).  This result is
qualitatively consistent with the location of the Break seen in
the ensemble Gaia DR3 vbroad catalog \citep[e.g., Figure 13
in][]{fremat2023}.

If the same criterion is used to define the Break versus effective
temperature, the Break spans from $6440 - 6600$ K.  Adopting the
color-temperature relations of \citet{PecautMamajek2013}, the color
limits from above imply limiting temperatures of $6470$ K and
$6670$ K, with interpolation.  These limits are 30 K and 70 K 
hotter than those inferred from spectroscopic temperature
measurements (mostly from Gaia DR3 spectra).  The offset is
consistent with the cooler temperatures inferred from  Gaia spectra
in this temperature range (41 K; Section 2).  With this offset in
mind and rounding to the nearest 50 K, we estimate that the Kraft
Break spans from $6450 - 6650$ K in temperature, with an uncertainty
of $\sim 50$ K on these limits.

\begin{figure}[ht!]
\centering
\plotone{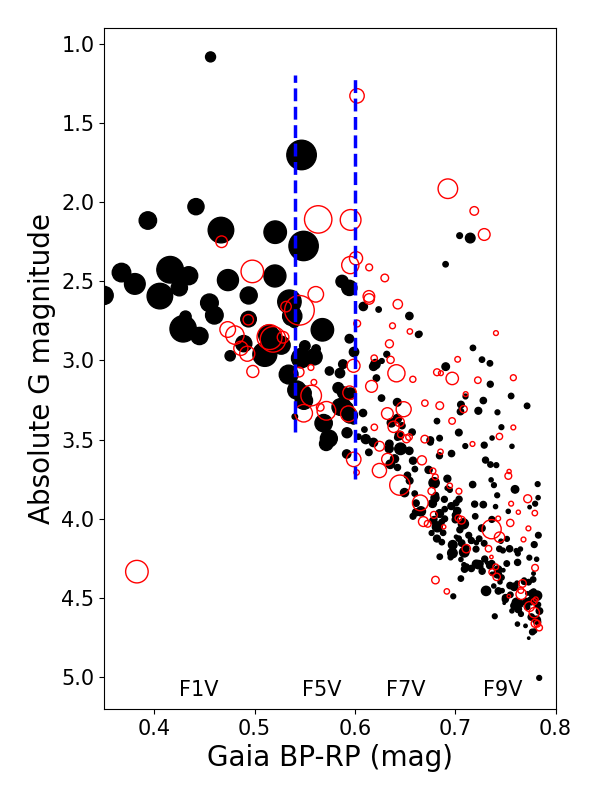}
\caption{The 33.33 pc sample of F stars is plotted on an absolute
$G$ versus $G_{BP}-G_{RP}$ diagram.  Young, evolved and
candidate binary stars (\textit{open circles}) are distinguished
from those that are not (\textit{filled circles}), as described 
in Section \ref{sec:exclude}.  Circle sizes are scaled by the
projected rotational velocity ($v$sin$i$) values.  The dashed
vertical lines mark the color boundaries of the Kraft Break.  The
anomolous outlier at the lower left of the plot is HD 194943 B; its
photometry may be biased by its brighter 1\farcs64 companion
\citep{makarov_fabricius2021}.
\label{fig:color_mag_break}}
\end{figure}

In Figure \ref{fig:color_mag_break} we illustrate the Kraft
Break on an absolute $G$ versus $G_{BP}-G_{RP}$ diagram, plotted
analogously to Figure 1 in \citet{Kraft1967}.  
The transition in rotation rates across the Break 
appears to hold over the $\sim 2$ magnitudes vertical
extent of the main sequence.

\subsection{Stellar Mass and Metallicity}

\begin{figure}[ht!]
\centering 
\plotone{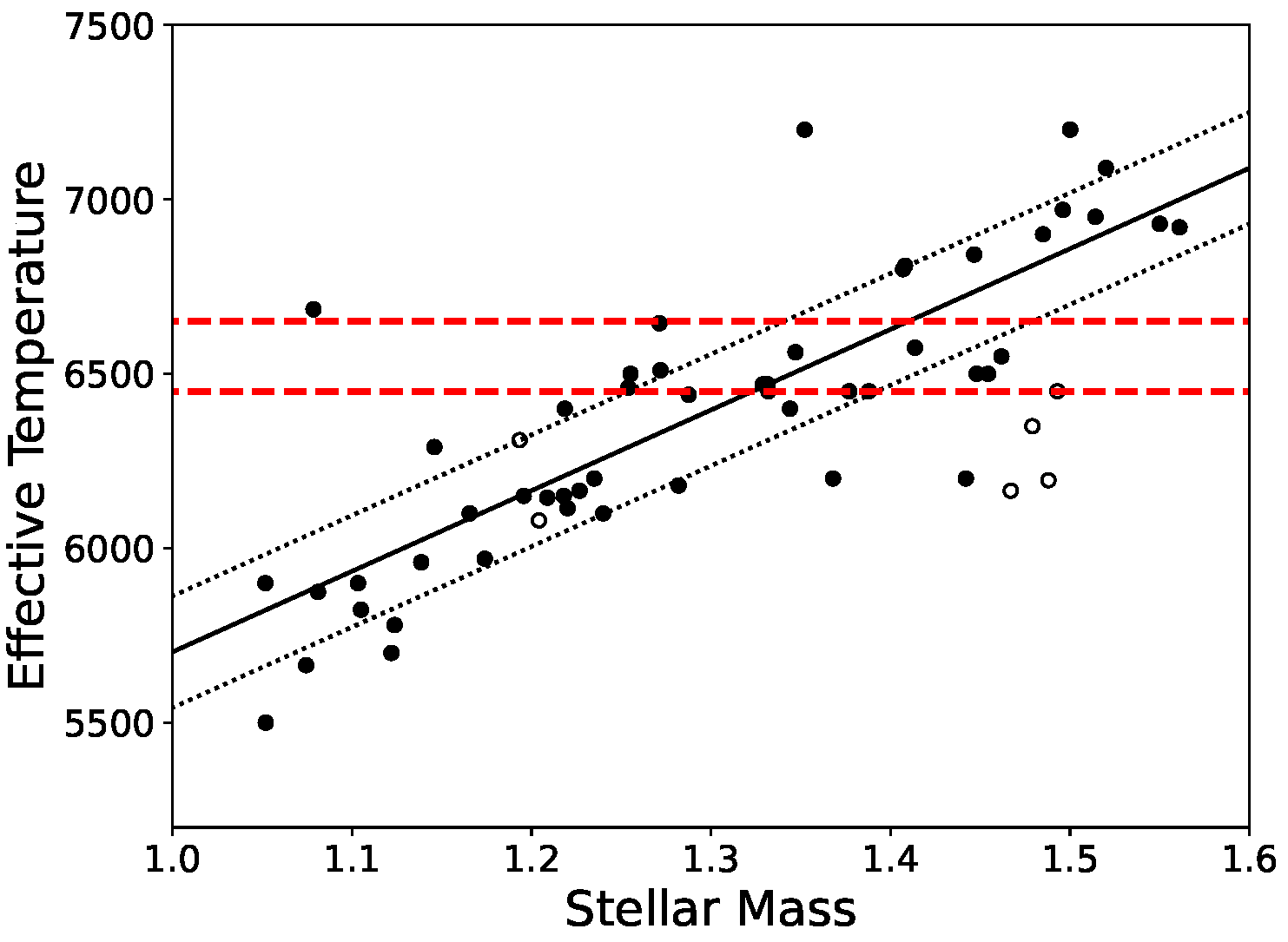}
    \caption{Effective temperature versus stellar mass for
    components of eclipsing binaries from \citet{torres2010}.  
    The temperature limits of the Kraft Break are shown as
    \textit{dashed lines}.  A linear fit to the eclipsing
    binary data is used define the corresponding mass limits
    of the Kraft Break (\textit{solid line}) and uncertainties
    (\textit{dotted lines}).  Components with log$g$ $< 4.0$
    dex (\textit{open circles}) are excluded from the fit.
    \label{fig:kraftbreak_masses}}
\end{figure}

We estimate the stellar mass limits of measured Kraft Break using
components of eclipsing binaries from \citet{torres2010}.  We
restrict the comparison sample to those with masses between 1.05
M$_\odot$ and 1.60 M$_\odot$; these stars have an average mass
uncertainty of 0.8\% and an average temperature uncertainty of 1.9\%.
From the sample of 59 stars within this mass range, we exclude 6 stars 
with log\,$g$ values below 4.0 dex as suspected of being evolved.
Using the Kraft Break temperature edges of 6450 K and 6650 K,
a linear fit to the mass versus temperature distribution of these
eclipsing
binary stars yields mass edges of 1.32 M$_\odot$ and 1.41 M$_\odot$
(Figure \ref{fig:kraftbreak_masses}).  We approximate a 1$\sigma$
uncertainty by varying
the y-intercept by $\pm$ 160 K to include 2/3rd of the data.  This
suggests the mass limits of the Break are uncertain by 0.07 M$_\odot$.  
These values are in good agreement using the color / temperature / mass
relations from \citet{PecautMamajek2013}; the $G_{BP}-G_{RP}$ color
edges of the Break correspond to masses of 1.31 M$_\odot$ and 1.39 
M$_\odot$, and the effective temperature edges of the Break correspond 
to 1.30 M$_\odot$ and 1.39 M$_\odot$.  We therefore adopt the mass
edges from the eclipsing binary analysis and estimate that the Kraft
Break spans from 1.32 M$_\odot$ to 1.41 M$_\odot$ in mass.

Metallicity affects the opacity within the outer, partial ionization
zones and thus can alter both the depth of convection and
convective turnover time.  These changes in turn affect the
efficiency of magnetic braking, with the general trend that 
metal-rich stars spin down more effectively than metal-poor stars
\citep[e.g.,][]{vansaders2019, amard_matt2020}.  The location
of the Kraft Break in color / temperature / mass may therefore be
metallicity dependent, but the 33.33 pc F star sample
studied here does not show any evidence of this.  The
Kraft Break is located within the same temperature / color
range for subsamples of stars above and below the median metallicity 
([Fe/H] = -0.43 dex); both subsamples span the full width of the
Break.  We caution however that the sample is too small to investigate
the effect of metallicity accurately, in part because the first-release
metallicity values from Gaia are considered to be biased
\citep{reinhold2023}; the median value is low compared to values
measured in detailed spectroscopic studies of F stars
\citep[e.g.,][]{Reddy2003}.

\section{The Kraft Break in the Hyades Open Cluster}  \label{sec:hyades}

The above results indicate that after a few hundred million years,
single, main sequence stars cooler than $\sim$ 6450 K will be slowly
rotating.  As a first check of when the Break is established, we
conduct a similar analysis of F stars in the well studied, nearby
\citep[46.75 $\pm$ 0.46 pc;][]{gaiacollaboration2017}, Hyades open
cluster, with an age of $650 \pm 50$ Myr \citep{perryman1998,
martin2018, gossage2018}.  Hyades members are selected from
\citet{brandner2023} and are restricted to those identified as
single and within the same $G_{BP}-G_{RP}$ color range
(0.327 - 0.784 mag).  This yielded 50 stars, 42 of which have
color, $v$sin$i$ and effective temperature measurements from
Gaia DR3.
We note that the average reddening for stars in the Hyades is
measured to be small, but possibly non-zero \citep[E(B$-$V)
$\le 0.001$ mag;][]{taylor2006}; we make no reddening corrections
to their colors.

\begin{figure*} [ht]
\centering
    \includegraphics[scale=0.4]{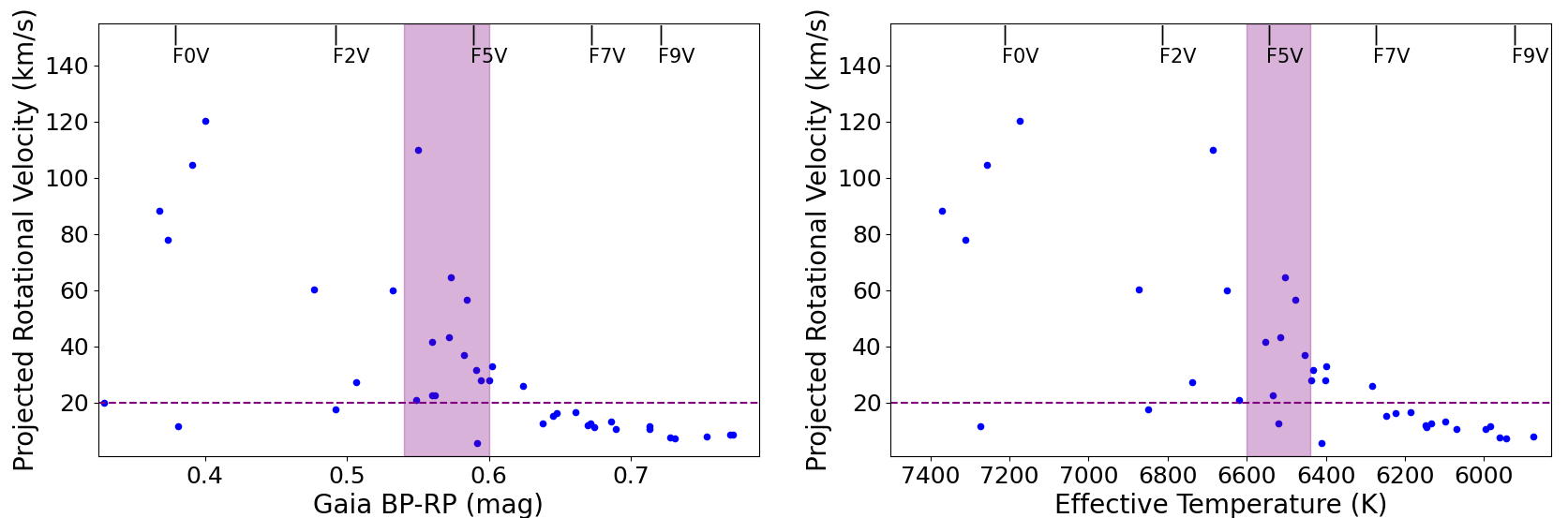}
    \caption{Shown are the projected rotational velocities versus 
    $G_{BP}-G_{RP}$ colors (\textit{left panel}) and effective
    temperatures (\textit{right panel}) of 49 single F stars in the
    Hyades.  The shaded regions define
    the location and width of the Kraft Break as determined from
    the 33.33 pc F star sample in Figure \ref{fig:kraftbreak_clean}.
    \label{fig:kraftbreak_hyades}}
\end{figure*}

Figure \ref{fig:kraftbreak_hyades} shows the projected rotational
distributions of these 42 Hyads versus $G_{BP}-G_{RP}$ color
and effective temperature.  The Break regions identified for the
33.33 pc sample are over-plotted for comparison.  The distributions
show a similar pattern, consistent with the initial investigation
by \citet{kraft1965}.  Stars hotter than the Break are consistent
with a randomly oriented, rapidly rotating sample, stars within the
Break region have an intermediate spread in $v$sin$i$ values,
and most of the stars cooler than the Break are slowly rotating
($v$sin$i$ $<$ 20 km/s).  The primary difference is that a few stars
close to but just redder / cooler than the Break have modest rotation.
Two stars that have colors redder than the Break region are identified
as rapidly rotating (with $G_{BP}-G_{RP}$ colors of 0.602 and 0.624 and
$v$sin$i$ values of 33.0 km/s and 26.1 km/s, respectively).  Five stars
that have temperatures cooler than the Break region are identified as
rapidly rotating (with temperatures spanning 6282 K to 6436 K, and
$v$sin$i$ values spanning 26.1 km/s - 33.0 km/s).  We speculate
that these mid-F stars
have not had time to slow sufficiently and establish the Kraft Break
as defined here.  Given these results along with evidence of how
rotation declines with age for stars less massive than the Break
\citep[e.g.,][]{gallet_bouvier2013,
vansaders2019, curtis2020, claytor2020, godoyrivera2021},
we speculate that the Break should be fully established by these
criterion by $1$ Gyr.

\section{Discussion and Conclusions} \label{sec:discussion}

The above analysis shows that the Kraft Break is well-defined, 
occurring at a central effective temperature of $\sim$ 6550 K,
central $G_{BP}-G_{RP}$ color of $\sim$ 0.57 mag, and central
mass of 1.37 M$_\odot$.  The Break is relatively sharp, having
a width of only 200 K in temperature, $\sim$ 0.06 mag in color
and $\sim$ 0.11 M$_\odot$ in mass.  We note that these values are
all larger than measurement uncertainties; the Break has width.
The location of the Kraft Break is consistent
with both theoretical \citep[e.g., Figure 6 in][]{vansaders_pinsonneault2013} 
and observational \citep[e.g.,][]{godoyrivera2021} values of 
the effective temperature and stellar mass at which stars exhibit a
marked change in their rotational evolution, but it is now more
sharply defined.  
This mass is consistent with the mass at which stars 
transition from having a thick convective envelope to having thin,
separate convective layers stemming from H and
He ionization \citep{cantiello2019, jermyn2022}.  
And this temperature closely matches the temperature
at which stars exhibit a dramatic shift in average surface gravity
values \citep[e.g., Figure 7 in][]{avallone2022}, confirming a
bona-fide change in structure.  Here we discuss some implications.

\subsection{Implications for Population Studies of Stars}

With very few exceptions, stars cooler than the Kraft Break that
are rapidly rotating are either young ($< 1$ Gyr), candidate
binary stars or evolved sub-giants.  Rotation thus provides a blunt
tracer for evolution or binarity in this mass regime, as is
generally known \citep[e.g.,][]{rampalli2023}.
Correspondingly, stars hotter than the Kraft Break with low $v$sin$i$ values
are likely pole-on rapid rotators \citep[e.g., Vega;][]{aufdenberg2006}.
This has implications for RV planet searches that target low $v$sin$i$
subsets of Sun-like stars to optimize RV precision
\citep[e.g.,][]{quinn2012, largrange2013}.
Assuming a low obliquity, planetary companions orbiting low $v$sin$i$
early-F stars will have reduced RV amplitudes (by sin$i$) and will
likely not transit.

Very rapidly rotating stars, those with equatorial velocities $\gtrsim 100$
km/s, with nearly pole-on orientations will also have biased apparent
colors, effective temperatures and luminosities because of their oblate
shape and gravity darkening \citep{owocki1994, vanbelle2012}.
As an example, \citet{che2011} show, based on interferometric imaging, 
that the nearly pole-on orientation of the F2 star
$\beta$ Cassiopeiae appears $\sim 400$ K hotter and $\sim 55$\%
more luminous than if viewed edge-on.  This will translate to a
spread in the single-star main sequence for stars more massive than
the Kraft Break.  If unrecognized, nearly pole-on rapid rotators could
be missclassified as photometric binaries, biasing comparisons with
evolutionary models \citep[e.g.,][]{brandner2023}.  This effect will 
also complicate age estimates that rely on a single main-sequence turnoff
point.  This is a known problem for nearby open clusters
\citep[e.g., Hyades, Praesepe, Ursa Majoris;][]{brandt_huang2015,
jones2015, gossage2019}, but it is likely a problem for any cluster
younger than $\sim 3$ Gyr, the approximate main sequence lifetime
of stars within the Kraft Break.

\subsection{Implications for Exoplanet Obliquities}

\citet{albrecht2022}, \citet{spalding_winn2022} and others have shown
that hot Jupiters orbiting cool stars tend to have low obliquities
while those orbiting hotter stars have a broad range of obliquities,
consistent with random inclinations.  These population
studies estimate a transition in the obliquity distribution at
$\sim$ 6100 K, corresponding to 1.17 $\pm 0.07$ M$_\odot$ from the
temperature / mass relation of Figure \ref{fig:kraftbreak_masses};
estimates from original work span from 6000 K to 6250 K, however 
\citep[e.g.,][]{schlaufman2010, winn2010}.  A favored interpretation
for the transition is
that the slower rotation and different internal structure of
cooler stars may enhance tidal realignment of the outer atmospheres
of stars \citep{dawson2014, rice2022}.
These cooler mass stars may have had a
similar, random distribution of obliquities initially, but for 
these stars tides induced by the planet
realigns the outer atmosphere of the star to rotate in the 
plane of its orbit.  This transition in the obliquity distribution
is often stated as occurring at the Kraft Break
\citep[e.g.,][]{rice2022, wright2023}, but the results
here suggest that it occurs some 450 K cooler than the Kraft Break.
The basic interpretation of tidal realignment may nevertheless be
correct.  Considering these phenomenon as a function of increasing
stellar mass and effective temperatures, tidal realignment appears
to become ineffective before magnetic
braking becomes ineffective.

\subsection{A Better Division Between Low Mass and Intermediate Mass Stars}

For both professional and pedagogical purposes, it is well accepted that 
``high mass stars" are those that end their life in a iron-core collapse
supernova.  Theoretical and observational studies find that this
occurs for stars more massive than $\sim$ $7 - 9$ M$_\odot$
\citep[see discussion in][]{cinquegrana2023}, with $8$ M$_\odot$
being a common approximate value.  Lower mass stars, those that end 
their life as a white dwarf, are similarly often divided into low mass
stars and intermediate mass stars, although the division between
these is less well-defined.

The most formal division between low mass stars and intermediate mass
stars stems from historical theoretical work.  \citet{iben1967}
first identified that stars more massive than 2.25 M$_\odot$, 
according to their calculations, do not experience a helium flash;
the helium cores of more massive stars are ignited during post-main
sequence evolution in a non-degenerate state.  Many stellar structure
and evolution textbooks now define this as a division between
low mass and intermediate mass stars \citep[e.g.,][]{lamerslevesque2017}.
And introductory astronomy textbooks adopt a similar division, 
rounded to ``about 2 solar masses" \citep[e.g.,][]{bennett2004}. Separating
stars by mass provides a way to discuss the differences in structure, 
changes in dominant fusion processes and rates of evolution between
Sun-like stars and
those more massive than the Sun, even though they all end their life
as white dwarfs.  But the specific reason for the division between low
mass and intermediate mass stars is usually not discussed
at the introductory level, possibly because it's too subtle to be 
relevant at this level.  But classroom experience confirms
that this omission, in contrast to the well-defined division between
intermediate mass and high mass stars, raises questions from astute
students.

The 'post-main sequence, non-degenerate core' division between low mass
stars and intermediate mass stars is not widely used by the 
scientific community, however.  In some cases ``intermediate mass" appears
to be used to simply distinguish a subsample of stars with larger or
smaller masses, but in many cases this division appears to be prompted
by the different stellar properties associated with stars above and below
the Kraft Break \citep[e.g.,][]{carlberg2014, pouilly2020,
nguyen2022, pereira2024, johnston2024}.

We propose that the Kraft Break provides a less ambiguous and a more useful 
division between low and intermediate mass stars.  The Break is
observationally well-defined, providing a clear distinction for main
sequence stars.  The Break is physically linked to a change in stellar
structure, the disappearance of the outer convective envelope.  This change
in structure is already the adopted transition point to ``intermediate mass
star" in some review studies \citep[e.g.][]{aerts2019}.  This division
could be related to topics that are discussed more thoroughly in both
introductory and graduate-level astronomy textbooks - the presence of
an outer convective zone.  And fortuitously the Kraft Break occurs
close to two other important stellar transition
points - the onset of core convection caused by the CNO-cycle
dominating energy production \citep{kippenhahn1990} and the red-edge of the
instability strip \citep{kaye1999, kurtz2022}.  For all of these reasons,
we propose that the Kraft
Break provides a more useful division, for both pedagogical and
scientific purposes, between what are classified as low mass stars
and intermediate mass stars.

\begin{acknowledgments}
We would like to thank Colin Kane, Becky Flores, Todd Henry and
John Meftah for helpful discussions, and we extend a special
thanks to Wei-Chun Jao, Jeremy Jones and Zachary Way for critically
reading a first draft.  
We thank Jamie Tayar for noting the alignment
of the Kraft Break with the global shift in stellar surface gravity 
values.  This research was supported by National
Science Foundation grant No. 2050829 to Georgia State University
under the Research Experiences for Undergraduates program.
This research has made use of the SIMBAD database, operated at CDS, Strasbourg, 
France.  This work has made use of data from the European Space Agency (ESA) 
mission {\it Gaia} (\url{https://www.cosmos.esa.int/gaia}), processed by the 
{\it Gaia} Data Processing and Analysis Consortium (DPAC, \url{https://www.cosmos.esa.int/web/gaia/dpac/consortium}).  Funding for the DPAC
has been provided by national institutions, in particular the institutions
participating in the {\it Gaia} Multilateral Agreement.  This research has made
use of NASA's Astrophysics Data System.
\end{acknowledgments}

\bibliography{main_bib}{}
\bibliographystyle{aasjournal}

\end{document}